\documentclass[%
amsmath,amssymb,
reprint,%
]{revtex4-1}
\usepackage{graphicx}
\usepackage{bm}
\usepackage[pdftex]{hyperref}
\usepackage[utf8]{inputenc}  

\newcommand{\ist}[1]{\overset{\footnotesize(\ref{#1})}{=}}
\newcommand{\circa}[1]{\overset{\footnotesize(\ref{#1})}{\approx}}
\newcommand{\soll}{\overset{\footnotesize !}{=}}
\newcommand{\iist}[2]{\overset{^{(\ref{#1})}}{\underset{^{(\ref{#2})}}{=}}}

\begin{document}


\title{Influence of gravitational waves on circular moving particles}
\author{Manfried Faber}
\affiliation{Atominstitut, Technical University, Vienna, Austria}
\author{Martin Suda}
\affiliation{Atominstitut, Technical University, Vienna, Austria}
\affiliation{AIT Austrian Institute of Technology GmbH, Digital Safety \&
Security, Security \& Communication Technologies,
Giefinggasse 4, 1210 Vienna, Austria}
\email{faber@kph.tuwien.ac.at}
\email{Martin.Suda@ait.ac.at}

\date{\today}

\begin{abstract}
  We investigate the influence of a gravitational wave background on particles in circular motion. We are especially interested in waves leading to stationary orbits. This consideration is limited to circular orbits perpendicular to the incidence direction. As a main result of our calculation we obtain in addition to the well-known alteration of the radial distance a time dependent correction term for the phase modifying the circular motion of the particle. A background of gravitational waves creates some kind of uncertainty.
\end{abstract}

\maketitle

\section{\label{sec:Int}Introduction}
One hundred years after Albert Einstein~\cite{Einstein:1916vd} predicted the existence of gravitational waves and after decades of searches, LIGO has succeeded to detect gravitational waves~\cite{2016PhRvLb}. Obviously, they originated from a merger of two stellar-mass black holes. With sizes of the interferometer arms of 4 km  LIGO is most sensitive in the frequency band 100–300~Hz. This success of LIGO could support the idea of the existence of gravitational waves in very different frequency ranges. In the Big Bang and in the later history of the universe a background of gravitational waves~\cite{Allen:1997ad} could have been produced, still today influencing elementary particles at the atomic scale, as one could speculate. The influence of such waves on particles on circular orbits seem not to have been treated yet. Especially interesting would be an influence on electrons in hydrogen. But apart from this special example the formalism could be valid quite generally for particles moving in a central force field.

The most important equation of physics of gravitational waves (GW) has been derived from general relativity (GR) by A. Einstein himself~\cite{Einstein:1916vd}. It is about radiation power or luminosity of GW and contains the quadrupole tensor. The components of this tensor has been calculated for standard problems by Misner, Thorne and Wheeler~\cite{Misner:1974qy}. The word-famous pulsar (PSR B1913-16) was the first one to be detected by Taylor and Hulse~\cite{Hulse:1974eb,Taylor:1979zz} where a GW-analysis has been executed. The two neutron stars are approaching each other exactly according to the prediction of GR caused by emission of GW. Important properties of this binary star system can be found in~\cite{Weisberg:2010zz,Weisberg:2016jye,Detweiler:1979wn}. The problem of spin precession has been developed further as well~\cite{Kramer:1998id}. The binary pulsar system PSR J0737-3039A has been investigated~\cite{Kramer:2006nb,Burgay:2003jj} confirming anew GR. Post-Newtonian approximative numerical procedures are used in order to execute computerized simulation of colliding black holes~\cite{Hannam:2007ik,Santamaria:2010yb,Baker:2005vv,Bruegmann:2003aw,Campanelli:2005dd}. Ultimatively, in September 14th 2015 the first GW (GW150914) from a merger of two black holes have been observed from LIGO~\cite{2016PhRvLa}. The direction where these GW have arrived could be approximately estimated~\cite{Connaughton:2016umz}. A second signal of GW (GW151226) has been observed too~\cite{2016PhRvLb}. In the meanwhile a numerical model of binary system of 2 stars between 40 and 100 solar masses has been discussed~\cite{Belczynski:2016obo}. A binary system of a 3 solar mass black hole with a X-ray star has been identified~\cite{Altamirano:2011aa} as well as spinning black holes~\cite{James:2015yla}. 

Trajectory and precession of spinning particles in external gravitational and electromagnetic fields have been treated analytically in ~\cite{Khriplovich:1997ni,Deriglazov:2015bqa,Deriglazov:2015wde,Deriglazov:2017jub}.

Tiny ripples in spacetime curvature propagate as waves with the speed of light and lead to periodic oscillations in the distance of test particles, as nicely expounded in many books and review articles, e.g. in \cite{Carroll,Cheng2005,schroeder2007,sharan2009,Braccini2016,Rebhan10}. An overview of detection of gravitational waves can be read in \cite{Roberto}. Therein it is as well described how gravitational waves arise from general relativity. In \cite{Pereira} a foundational review about gravitational waves is given and a critical review of the standard linear approach of the theory is depicted.

The aim of this paper is to investigate particles moving in a central force field under the influence of gravitational waves. To achieve this goal we have to modify the equation of geodesic deviation including additional force terms. We are especially interested in uncertainties of the orbits and conditions for stationary orbits.

The paper is organized as follows: In chapter II we repeat the results obtained for distances of test particles in the presence of plane gravitational waves. Afterwards, in chapter III, we discuss an extension of the problem by considering the influence of such gravitational waves on circular moving particles. In chapter IV results of the previous chapter are graphically displayed and discussed. Finally in chapter V conclusions are drawn.

\section{\label{sec:part} Motion of test particles}
In order to obtain a coordinate independent measure of the wave's influence, the relative motion of two nearby particles can be considered, see e.g. in \cite{Carroll}. It can be described by the geodesic equation in four-dimensional space-time $x^\mu=(ct,x,y,z)$. The four-velocity is given by  $U^\mu=\frac{dx^\mu}{d\tau}$ and the distance vector $S^\mu$ is a solution of the differential equation
\begin{eqnarray}\label{GG}
\frac{D^2}{d\tau^2}S^\mu=\gamma^2\partial_0^2S^\mu=S^\sigma U^{\nu}U^{\rho}{R^\mu}_{\nu\rho\sigma}
\end{eqnarray}
with the Minkowski metric $\eta^{\mu\lambda}=\mathrm{diag}(-1,1,1,1)$ in the flat background. The double differential $\frac{D^2}{d\tau^2}$ points to the parallel transport in general relativity and the differential of the proper time is given by $d\tau=\frac{1}{\gamma}dt$ with $\gamma=\frac{1}{\sqrt{1-\beta^2}}$ and $\beta=\frac{v}{c}$.

In curved space-time the linearized Riemann-tensor $R_{\mu\nu\rho\sigma}$ is given by
\begin{eqnarray}\begin{aligned}\label{RT}
R_{\mu\nu\rho\sigma}=\frac{1}{2}(&\partial_\rho\partial_\nu h_{\mu\sigma}
+\partial_\sigma\partial_\mu h_{\nu\rho}\\
&-\partial_{\sigma}\partial_\nu h_{\mu\rho}-\partial_\rho\partial_\mu h_{\nu\sigma}).
\end{aligned}\end{eqnarray}
Here the metric $g_{\mu\nu}$ has been approximated by $g_{\mu\nu}=\eta_{\mu\nu}+h_{\mu\nu}$ with $|h_{\mu\nu}|\ll{1}$.

For a gravitational wave $h_{\mu\nu}$, propating in $z-$direction with the velocity of light $c$, one obtains harmonic oscillations, which neglecting additional phases may be written as (the real part of) plane waves \cite{Carroll,Cheng2005,schroeder2007,sharan2009,Braccini2016,Rebhan10}
\begin{eqnarray}\label{GW}
h_{\mu\nu}=C_{\mu\nu}\cos(kz-\omega_gt)
\end{eqnarray}
where $\omega_g=ck$ is the frequency of the gravitational wave ($k$ is the wave number). The constant quantities $C_{\mu\nu}\ll 1$ form a symmetric $(0,2)$ tensor with $C_{0\nu}=C_{3\nu}=0$ and $C_{22}=-C_{11}$ as well as $C_{12}=C_{21}$. They generate a time-varying quadrupole deformation. The two parameters $C_{11}$ and $C_{12}$ are sufficient to describe any quadrupole deformation in the xy-plane.

\begin{figure}[h!]
\includegraphics[scale=0.60]{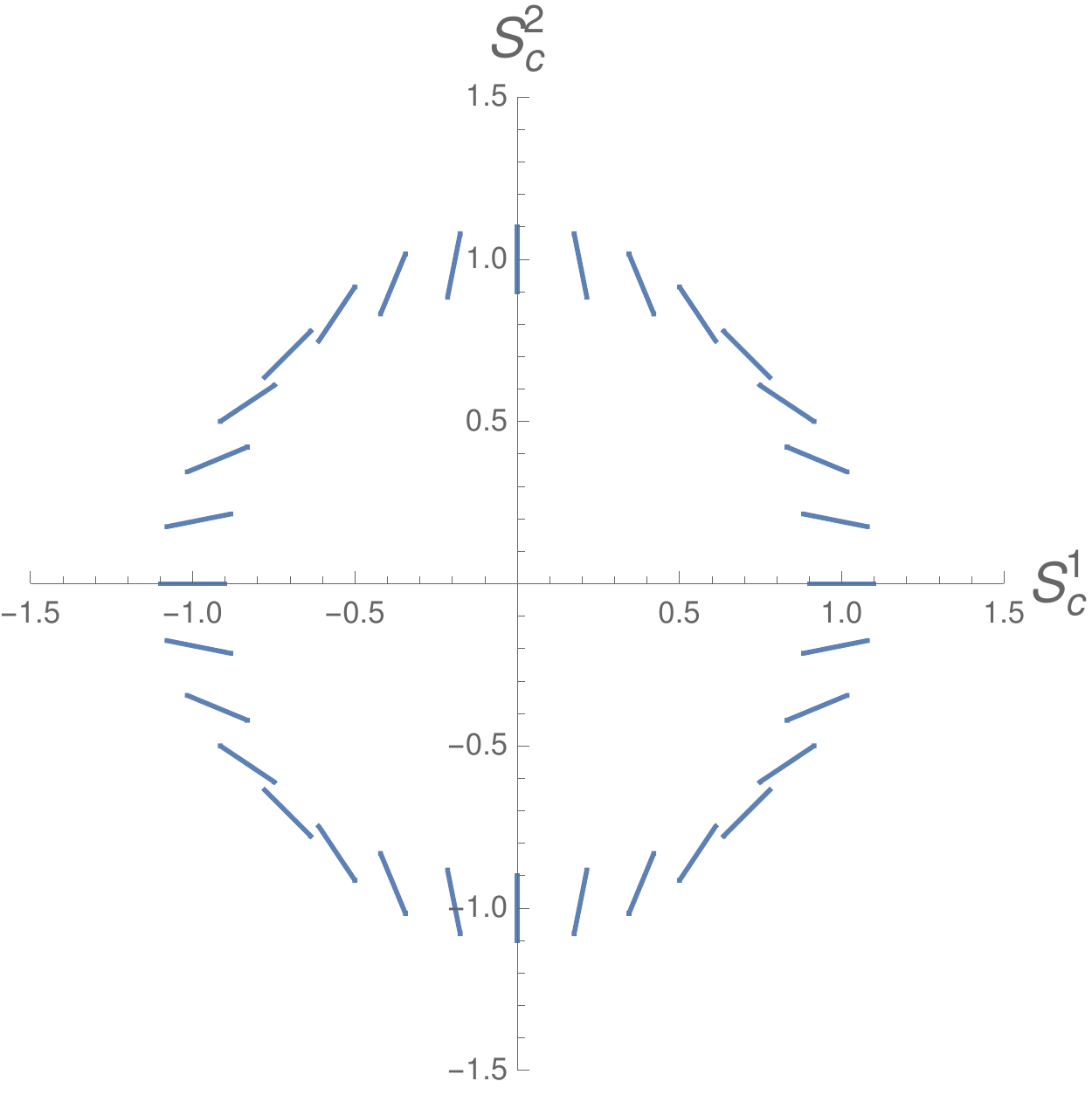}\caption{Linear oscillations of $(S_c^1(t),S_c^2(t))$ with time $t$ for various values of $S_c^1(0)$ and $S_c^2(0)$ with $\sqrt{[S_c^1(0)]^2+[S_c^2(0)]^2}=1$ and $C_{11}=0.2$.}
\label{BS1CBS2C}
\end{figure}

Because the Riemann-tensor is first order, the corrections to $U^{\nu}$ may be ignored, and, for slowly moving particles ($\tau=x^{0}=ct$), we have $U^{\nu}=(1,0,0,0)$, setting $c=1$. As a consequence $R_{\mu00\sigma}=\frac{1}{2}\partial_0^2h_{\mu\sigma}$ and the geodesic equation~\eqref{GG} becomes
\begin{eqnarray}\label{GE}
    \frac{\partial^2}{\partial t^2}S^\mu
    \ist{GG}S^\sigma\frac{\partial^2}{\partial t^2}\frac{{h^\mu}_\sigma}{2}.
\end{eqnarray}
$C_{11}$ and $C_{12}$ differ only by a $\pi/4$-rotation in the xy-plane. Therefore we can choose without loss of generality $C_{12}=0$ resulting in
\begin{eqnarray}\label{C11}
\begin{aligned}
&\partial_0^2S^1\ist{GE}S^1\partial_0^2\frac{h_{11}}{2},\quad 
&&&\partial_0^2S^0\ist{GE}0,\\
&\partial_0^2S^2\ist{GE}-S^2\partial_0^2\frac{h_{11}}{2},\quad
 &&&\partial_0^2S^3\ist{GE}0,
\end{aligned}\end{eqnarray}
with $h_{11}$ given by Eq.~(\ref{GW}). For $C_{11}\ll 1$ the solutions of these equations are  given in refs.~\cite{Carroll,Cheng2005,schroeder2007,sharan2009,Braccini2016,Rebhan10} by
\begin{eqnarray}\begin{aligned}\label{C11L}
&S^1\approx S_c^1(t):=S_c^1(0)\left[1+\frac{h_{11}}{2}\right],\\
&S^2\approx S_c^2(t):=S_c^2(0)\left[1-\frac{h_{11}}{2}\right],\\
&S^0:=S^3:=0.
\end{aligned}\end{eqnarray}
$S_c^1(t)$ and $S_c^2(t)$ are the coordinates of the separation vector in the xy-plane. The oscillations in x and y coordinates are $180^\circ$ out of phase and lead therefore to linear oscillations around the vector $(S_c^1(0),S_c^2(0))$, as shown in Fig.~\ref{BS1CBS2C}.
To increase the visibility we have chosen an unrealistic large value for $C_{11}$. The functions in Eq.~(\ref{C11L}) fulfil Eq.~(\ref{C11}) up to order $C_{11}^2$ only\footnote{With the approximation symbol ``$\approx$'' we indicate accuracy up to order $C_{11}$.}
\begin{eqnarray}\begin{aligned}\label{genau}
S_c^1(0)&(-\omega_g^2)\frac{h_{11}}{2}
\circa{C11}S_c^1(t)(-\omega_g^2)\frac{h_{11}}{2},\\
S_c^2(0)&\omega_g^2\,\frac{h_{11}}{2}
\circa{C11}-S_c^2(t)(-\omega_g^2)\frac{h_{11}}{2},
\end{aligned}\end{eqnarray}

\section{\label{sec:Bewwinkel}Circular motion with  constant angular velocity}
Let us consider a light particle rotating in the xy-plane on a circle with radius $r$ with constant angular velocity $\omega$ around a heavy particle, like in Bohr's model for the hydrogen atom. We assume that the radius $r$ and the azimuthal angle $\varphi=\omega t$ of this motion are slightly disturbed by a gravitational wave with ``+''-polarisation incident in 3-direction with $h_{\mu\nu}$ given by Eq.~\ref{GW} and the symmetric tensor
\begin{eqnarray}\label{cmn}
C_{0\nu}=C_{12}=C_{3\nu}=0,\quad C_{22}=-C_{11}.
\end{eqnarray}

In a generalised equation of motion we have to take into account the gravitational force, the rhs. of Eq.~(\ref{GG}), and the central force keeping the light particle on a circle. For the assumed circular motion it can be approximated by the expression for the centripetal force
\begin{eqnarray}\label{BG}
\partial_0^2S^\mu=S^\sigma\frac{U^{\nu}}{\gamma}\frac{U^{\rho}}{\gamma}
{R^\mu}_{\nu\rho\sigma}+(0,\vec a_r,0)^\mu,
\end{eqnarray}
where $\vec a_r$ is the radial acceleration in the xy-plane caused by the external centripetal force.

The geometrical term $S^\sigma U^{\nu}U^{\rho}{R^\mu}_{\nu\rho\sigma}$ representing the curved space-time due to gravitational waves simplifies drastically due to our special assumptions~(\ref{GW}) and (\ref{cmn}) --- wave propagating in z-direction and particle rotating in  the xy-plane. $h_{\mu\nu}$ of Eq.~(\ref{GW}) depends only on $z$ and $t$. In the curvature tensor $R_{\mu\nu\rho\sigma}$ of Eq.~(\ref{RT}) the derivatives of $h_{\mu\nu}$ are always contracted either with $S^\mu$ or $U^\mu$. Due to $S^3=U^3=0$ finally only time derivatives of $h_{11}$ and $h_{22}=-h_{11}$ matter. Therefore, the contributions of the geometrical term read
\begin{eqnarray}\begin{aligned}\label{gT}
S^\sigma&U^{\nu}U^{\rho}{R^0}_{\nu\rho\sigma}\ist{RT}\\
&=-\partial_0^2\frac{h_{11}}{2}[S^0(U^1U^1-U^2U^2)-S^1U^0U^1+S^2U^0U^2],\\
S^\sigma&U^{\nu}U^{\rho}{R^1}_{\nu\rho\sigma}\ist{RT}\\
=&\partial_0^2\frac{h_{11}}{2}[-S^0U^0U^1+S^1U^0U^0],\\
S^\sigma&U^{\nu}U^{\rho}{R^2}_{\nu\rho\sigma}\ist{RT}\\
&=\partial_0^2\frac{h_{11}}{2}[S^0U^0U^2-S^2U^0U^0],
\end{aligned}\end{eqnarray}
As expected, due to the factor
\begin{eqnarray}\label{AblWel}
\partial_0^2\frac{h_{11}}{2}=-\frac{C_{11}}{2}\,\omega_g^2\,\cos(kz-\omega_gt).
\end{eqnarray}
all these contributions are of first order in $C_{11}$.

In the spirit of the approximate solution~(\ref{C11L}) we expect that the distance vector of pure circular motion indicated by the subscript $0$
\begin{eqnarray}\label{S0}
S_0^\mu=(0,r\cos\varphi,r\sin\varphi,0),\quad\varphi(t)=\omega t,
\end{eqnarray}
is modified by terms linear in $C_{11}$. The modifications of the radial distance according to Eq.~(\ref{C11}) lead to Coriolis forces inducing variations of the angular velocity. Therefore, we try the ansatz
\begin{align}
&\;S^0:=S^0(t),\label{ansatz0}\\
&\begin{aligned}S^1:=&R_1(t)\cos\varphi_1(t)=\label{ansatz1}\\
:=&r\,[\,1+\frac{h_{11}(t)}{2}\,]\cos[\varphi(t)-\eta_1(t)],\end{aligned}\\
&\begin{aligned}S^2:=&R_2(t)\sin\varphi_2(t)=\label{ansatz2}\\
:=&r\,[\,1-\frac{h_{11}(t)}{2}\,]\sin[\varphi(t)+\eta_2(t)],\end{aligned}\\
&S^3:=0,\label{ansatz3}
\end{align}
where $S^0$, $\eta_1$ and $\eta_2$ are of first order in $C_{11}$.
For the velocity vector $U^\mu$ we get
\begin{align}
&\frac{U^0}{\gamma}=1+\partial_0S^0(t),\label{geschw0}\\
&\frac{U^1}{\gamma}=\partial_0S^1\ist{ansatz1}\partial_0R_1\cos\varphi_1
-R_1\sin\varphi_1\,\dot\varphi_1,\label{geschw1}\\
&\frac{U^2}{\gamma}=\partial_0S^2\ist{ansatz2}\partial_0R_2\sin\varphi_2
+R_2\cos\varphi_2\,\dot\varphi_2,\label{geschw2}\\
&U^3=\partial_0S^3\ist{ansatz3}0\label{geschw3}
\end{align}
using dots for time-derivatives $\partial_0$. With Eqs.~(\ref{ansatz0})-(\ref{geschw3}) we are ready to calculate the geometrical terms~(\ref{gT}) up to order $C_{11}$
\begin{align}
&\begin{aligned}\label{gT0}
S^\sigma&\frac{U^{\nu}}{\gamma}\frac{U^{\rho}}{\gamma}{R^0}_{\nu\rho\sigma}\circa{gT}\\
&\approx-\partial_0^2\frac{h_{11}}{2}[-S_0^1\partial_0S_0^1+S_0^2\partial_0S_0^2]
\circa{S0}\\&\approx-\partial_0^2\frac{h_{11}}{2}\,r^2\omega\sin(2\varphi)
\ist{AblWel}\\
&\ist{S0}\frac{C_{11}}{2}\,\omega_g^2\,\cos(kz-\omega_gt)
\,r^2\omega\sin(2\omega t),\end{aligned}\\
&\begin{aligned}\label{gT1}
S^\sigma&\frac{U^{\nu}}{\gamma}\frac{U^{\rho}}{\gamma}{R^1}_{\nu\rho\sigma}\circa{gT}
\partial_0^2\frac{h_{11}}{2}S^1\ist{ansatz1}\\
&=\partial_0^2R_1\cos\varphi_1,\end{aligned}\\
&\begin{aligned}\label{gT2}
S^\sigma&\frac{U^{\nu}}{\gamma}\frac{U^{\rho}}{\gamma}{R^2}_{\nu\rho\sigma}\circa{gT}
-\partial_0^2\frac{h_{11}}{2}S^2\ist{ansatz2}\\
&=\partial_0^2R_2\sin\varphi_2,\end{aligned}
\end{align}

Thus, the equation of motion~(\ref{BG}) for the time-component reads 
\begin{eqnarray}\begin{aligned}\label{BG0}
\partial_0^2S^0&\circa{gT0}\frac{C_{11}}{4}\,r^2\omega\,\omega_g^2\\
&\left\{\sin[(2\omega-\omega_g)t+kz]+\sin[(2\omega+\omega_g)t-kz]\right\}
\end{aligned}\end{eqnarray}
with a solution of order $C_{11}$
\begin{eqnarray}\begin{aligned}\label{Sol0}
S^0&\circa{BG0}-\frac{C_{11}}{4}\,r^2\omega\,\omega_g^2\\
&\left\{\frac{\sin[(2\omega-\omega_g)t+kz]}{(2\omega-\omega_g)^2}
+\frac{\sin[(2\omega+\omega_g)t-kz]}{(2\omega+\omega_g)^2}\right\}.
\end{aligned}\end{eqnarray}

\begin{figure}[h!]
\includegraphics[scale=1.0]{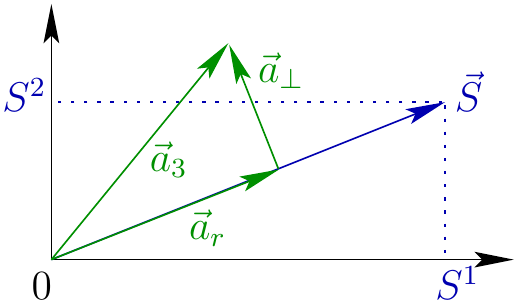}
\caption{Decomposition of the acceleration $\vec a_3$ in components parallel and perpendicular to the distance vector $\vec S$ in the xy-plane.}
\label{compForce}
\end{figure}

More terms we get for the x- and y-components
\begin{align}\begin{aligned}\label{DQ1}
\partial_0^2S^1&\ist{ansatz1}\partial_0^2R_1\cos\varphi_1
-2\partial_0R_1\sin\varphi_1\dot\varphi_1-\\&-R_1\cos\varphi_1\dot\varphi_1^2
-R_1\sin\varphi_1\ddot\varphi_1,\end{aligned}\\
\begin{aligned}\label{DQ2}\partial_0^2S^2&\ist{ansatz2}
\partial_0^2R_2\sin\varphi_2
+2\partial_0R_2\cos\varphi_2\dot\varphi_2-\\&-R_2\sin\varphi_2\dot\varphi_2^2
+R_2\cos\varphi_2\ddot\varphi_2.\end{aligned}
\end{align}
To simplify the notation for the four terms on the rhs we use two-dimensional vectors in the xy-plane. Further on we denote them by $\vec a_1,\,\vec a_2,\,\vec a_3$ and $\vec a_4$ and use dots for time derivatives $\partial_0$. We realise that the first terms $\vec a_1$ in these two equations are the geometrical terms~(\ref{gT1}) and (\ref{gT2}). The third terms
\begin{eqnarray}\label{a3}
\vec a_3:=-\begin{pmatrix}R_1\cos\varphi_1\,\dot\varphi_1^2\\
R_2\sin\varphi_2\,\dot\varphi_2^2\end{pmatrix}\iist{ansatz1}{ansatz2}
-\begin{pmatrix}S_1\,\dot\varphi_1^2\\S_2\,\dot\varphi_2^2\end{pmatrix}
\end{eqnarray}
include the centrifugal acceleration $\vec a_r$ acting in radial direction $\vec S=\bigl(\begin{smallmatrix}S^1\\S^2\end{smallmatrix}\bigr)$, see Fig.~\ref{compForce},
\begin{eqnarray}\label{ar}
\vec a_r:=\frac{\vec a_3\vec S}{{\vec S}^2}\,\vec S
=-\frac{(S^1)^2\dot\varphi_1^2+(S^2)^2\dot\varphi_2^2}{(S^1)^2+(S^2)^2}\,\vec S.
\end{eqnarray}
Inserting
\begin{eqnarray}\begin{aligned}\label{punktphi}
\dot\varphi_1^2\ist{ansatz1}(\dot\varphi-\dot\eta_1)^2\circa{S0}
\omega^2-2\omega\dot\eta_1,\\
\dot\varphi_2^2\ist{ansatz1}(\dot\varphi+\dot\eta_2)^2\circa{S0}
\omega^2+2\omega\dot\eta_2,
\end{aligned}\end{eqnarray}
we get
\begin{eqnarray}\begin{aligned}\label{azb}
\vec a_r\ist{ar}&-\omega^2\vec S\left[1-\frac{2}{\omega}
\frac{\dot\eta_1(S^1)^2-\dot\eta_2(S^2)^2}{{\vec S}^2}\right]=\\
\iist{ansatz1}{ansatz2}&-\omega^2\vec S
\left[1-\frac{2}{\omega}\left(
\dot\eta_1\cos^2\varphi-\dot\eta_2\sin^2\varphi\right)\right].
\end{aligned}\end{eqnarray}
The remainder of $\vec a_3$
\begin{eqnarray}\begin{aligned}\label{azim}
\vec a_\perp:=&\vec a_3-\vec a_r\iist{a3}{azb}\\=&2\omega
\begin{pmatrix}S^1[\dot\eta_1-\dot\eta_1\cos^2\varphi+\dot\eta_2\sin^2\varphi]\\
S^2[-\dot\eta_2-\dot\eta_1\cos^2\varphi+\dot\eta_2\sin^2\varphi]\end{pmatrix}\\
\iist{ansatz1}{ansatz2}&2\omega
\begin{pmatrix}S^2(\dot\eta_1+\dot\eta_2)\sin\varphi\cos\varphi\\
-S^1(\dot\eta_1+\dot\eta_2)\sin\varphi\cos\varphi\end{pmatrix}
\end{aligned}\end{eqnarray}
contributes to the azimuthal accelerations together with the second and forth term in Eqs.~(\ref{DQ1}) and (\ref{DQ2}). Since there are no forces leading to azimuthal accelerations we have to set their sum to zero
\begin{eqnarray}\label{aphi0}
\underbrace{\vec a_3-\vec a_r}_{\vec a_\perp}+\vec a_2+\vec a_4\soll0.
\end{eqnarray}
This is a condition for the time dependence of the angular velocity. It leads to
\begin{eqnarray}
\omega\sin(2\varphi)(\dot\eta_1+\dot\eta_2)-\omega\,\partial_0h_{11}
+\ddot\eta_1=0,\label{aphi1}\\
-\omega\sin(2\varphi)(\dot\eta_1+\dot\eta_2)-\omega\,\partial_0h_{11}
+\ddot\eta_2=0.\label{aphi2}
\end{eqnarray}
The sum of these two equations
\begin{eqnarray}\label{dd1+2}
\ddot\eta_1+\ddot\eta_2\iist{aphi1}{aphi2}2\omega\,\partial_0h_{11}
\end{eqnarray}
can be solved by
\begin{eqnarray}\label{d1+2}
\dot\eta_1+\dot\eta_2\ist{dd1+2}2\omega\,h_{11}
\end{eqnarray}
and by
\begin{eqnarray}\label{1+2}
\eta_1+\eta_2\ist{GW}-2\,C_{11}\frac{\omega}{\omega_g}\sin(kz-\omega_gt).
\end{eqnarray}
The difference of the two equations~(\ref{aphi1}) and (\ref{aphi2})
\begin{eqnarray}\begin{aligned}\label{dd2-1}
\ddot\eta_2-\ddot\eta_1\iist{aphi1}{aphi2}
&2\omega\sin(2\varphi)(\dot\eta_1+\dot\eta_2)
\ist{d1+2}4\omega^2\sin(2\varphi)\,h_{11}\ist{GW}\\
=&\,4\,C_{11}\,\omega^2\sin(2\varphi)\cos(kz-\omega_gt)
\end{aligned}\end{eqnarray}
can be solved by
\begin{eqnarray}\begin{aligned}\label{2-1}
&\eta_2-\eta_1\ist{dd2-1}2\,C_{11}\,\omega^2\\
&\left\{\frac{\sin[(2\omega-\omega_g)t+kz]}{(2\omega-\omega_g)^2}+
\frac{\sin[(2\omega+\omega_g)t-kz]}{(2\omega+\omega_g)^2}\right\}.
\end{aligned}\end{eqnarray}
We get the results
\begin{eqnarray}\label{eta12}
&&\eta_{1,2}\iist{1+2}{2-1}-C_{11}\,
\left\{\frac{\omega}{\omega_g}\sin(kz-\omega_gt)\right.\\
&&\left.\pm\omega^2\left[\frac{\sin[(2\omega-\omega_g)t+kz]}
{(2\omega-\omega_g)^2}+\frac{\sin[(2\omega+\omega_g)t-kz]}
{(2\omega+\omega_g)^2}\right]\right\}=\nonumber\\
&&\ist{Sol0}\frac{\omega}{\omega_g}\left\{-C_{11}\sin(kz-\omega_gt)
\pm\frac{4S^0}{\omega_gr^2}\right\},\nonumber
\end{eqnarray}
where according to Eq.~(\ref{Sol0}) $S^0$ is of order $C_{11}$.\\[5mm]\noindent
Result: We have obtained the time-dependent correction terms $\eta_{1,2}=\eta_{1,2}(t)$ which have to be subtracted from (added to) $\varphi(t)=\omega t$ in Eqs.~(\ref{ansatz1}) and (\ref{ansatz2}) when circular moving particles are influenced by gravitational waves, Eq.~(\ref{GW}). We have found a solution $S^\mu$ according to the ansatz~(\ref{ansatz0})-(\ref{ansatz3}) fullfilling the equation of motion~(\ref{BG}) up to first order in $C_{11}$.

The distance vector $\vec S=(S^1,S^2)$ from the first to the second particle, shown in Fig.~\ref{compForce}, can be represented by the polar coordinates
\begin{eqnarray}\begin{aligned}\label{PolKoo}
R:=\sqrt{(S^1)^2+(S^2)^2},\quad\phi:=\arctan\frac{S^2}{S^1}
\end{aligned}\end{eqnarray}
with $S^1$ and $S^2$ given in Eqs.~(\ref{ansatz1}) and (\ref{ansatz2}).

The gravitational wave modifies the radius $R$ of the circular motion and in consequence the angular velocity deviates from $\omega$.

\section{\label{sec:C}Results and discussion}
\begin{figure}[h!]
\includegraphics[scale=0.50]{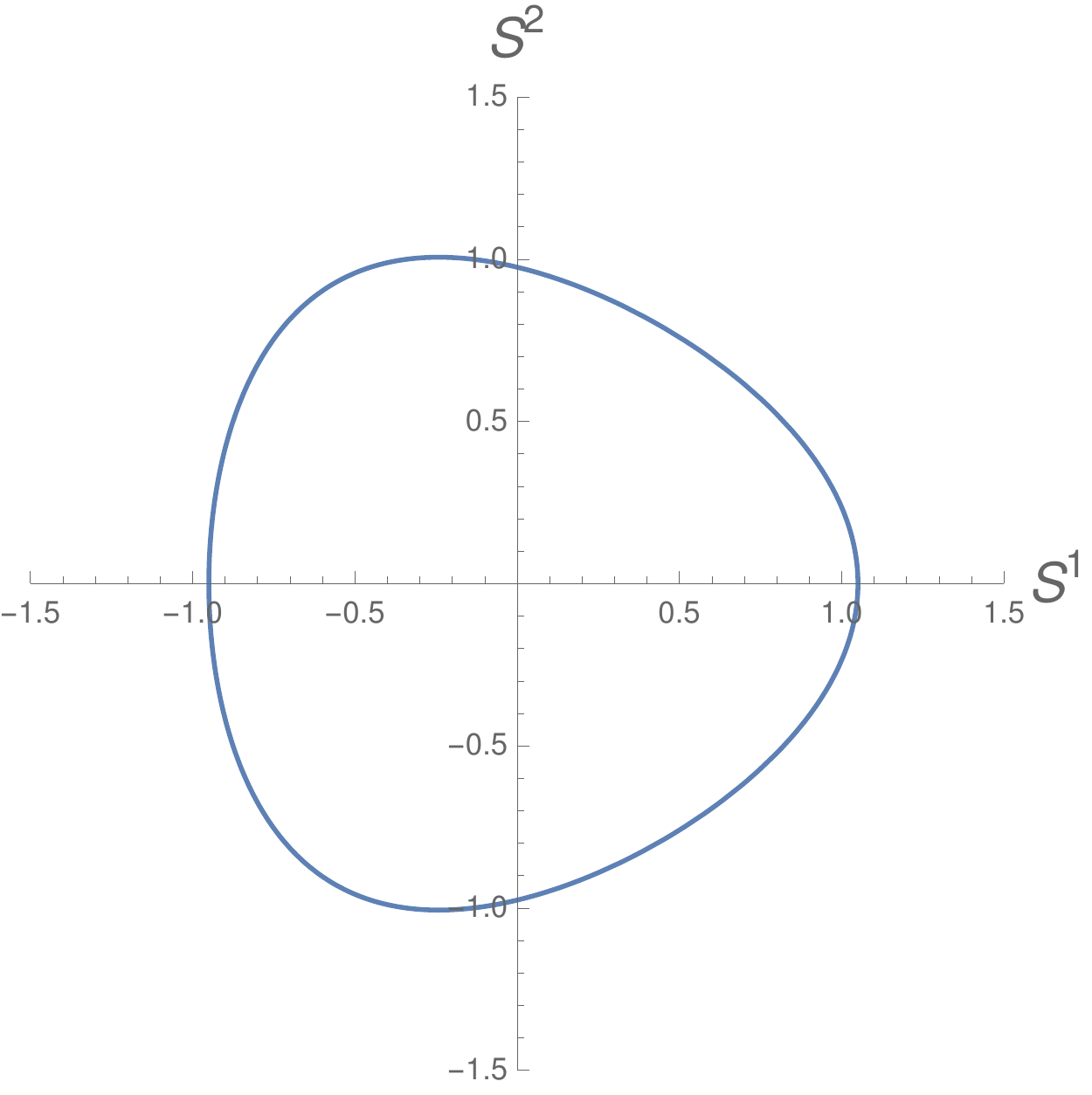}
\caption{$\frac{C_{11}}{2}=0.05$, $kz=0$, $\frac{\omega_g}{\omega}=1$, $t=T$.}
\label{BS2}
\end{figure}

\begin{figure}[h!]
\includegraphics[scale=0.50]{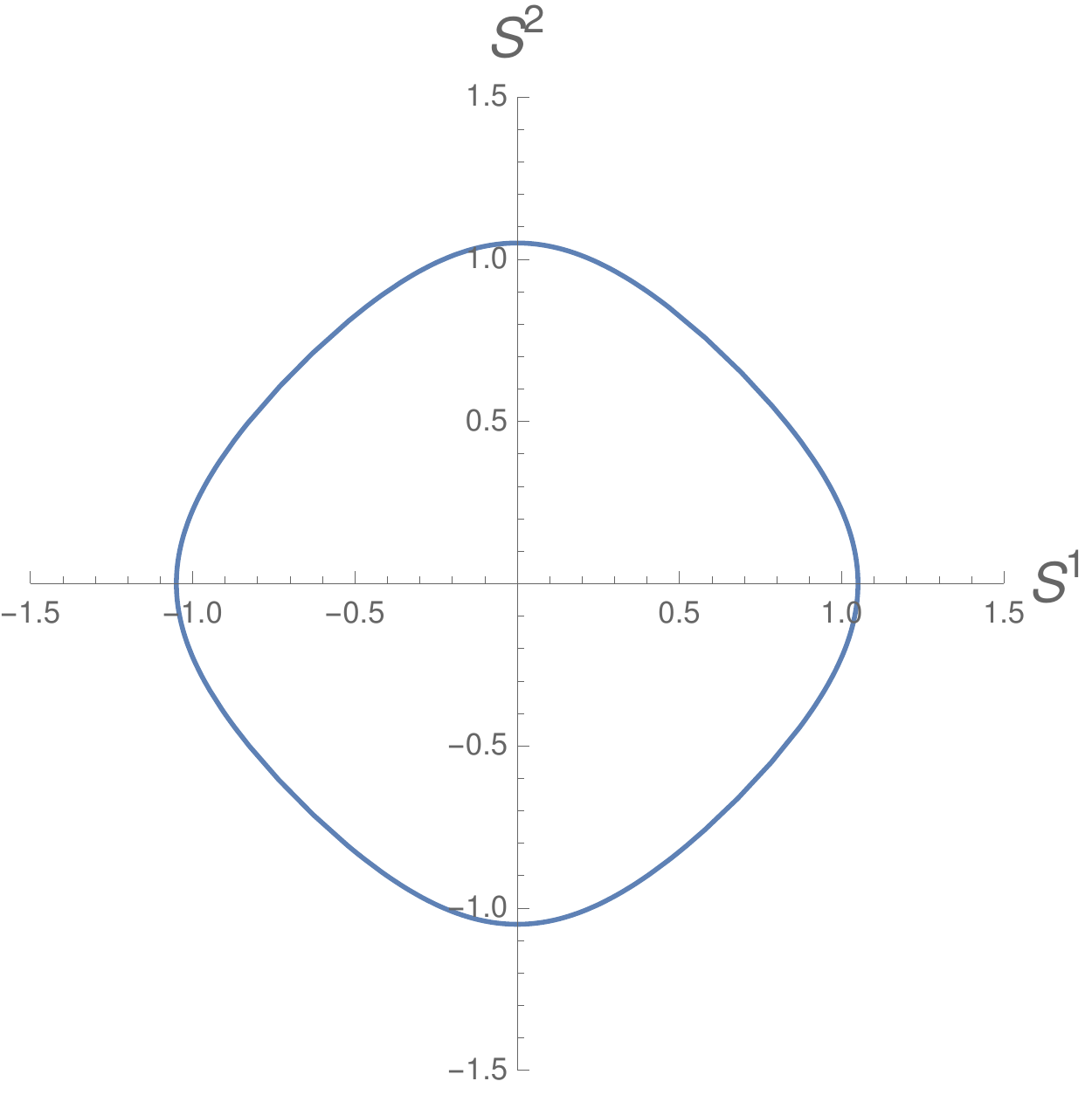}
\caption{$\frac{C_{11}}{2}=0.05$, $kz=0$, $\frac{\omega_g}{\omega}=2$, $t=T$.}
\label{BS3}
\end{figure}

\begin{figure}[h!]
\includegraphics[scale=0.50]{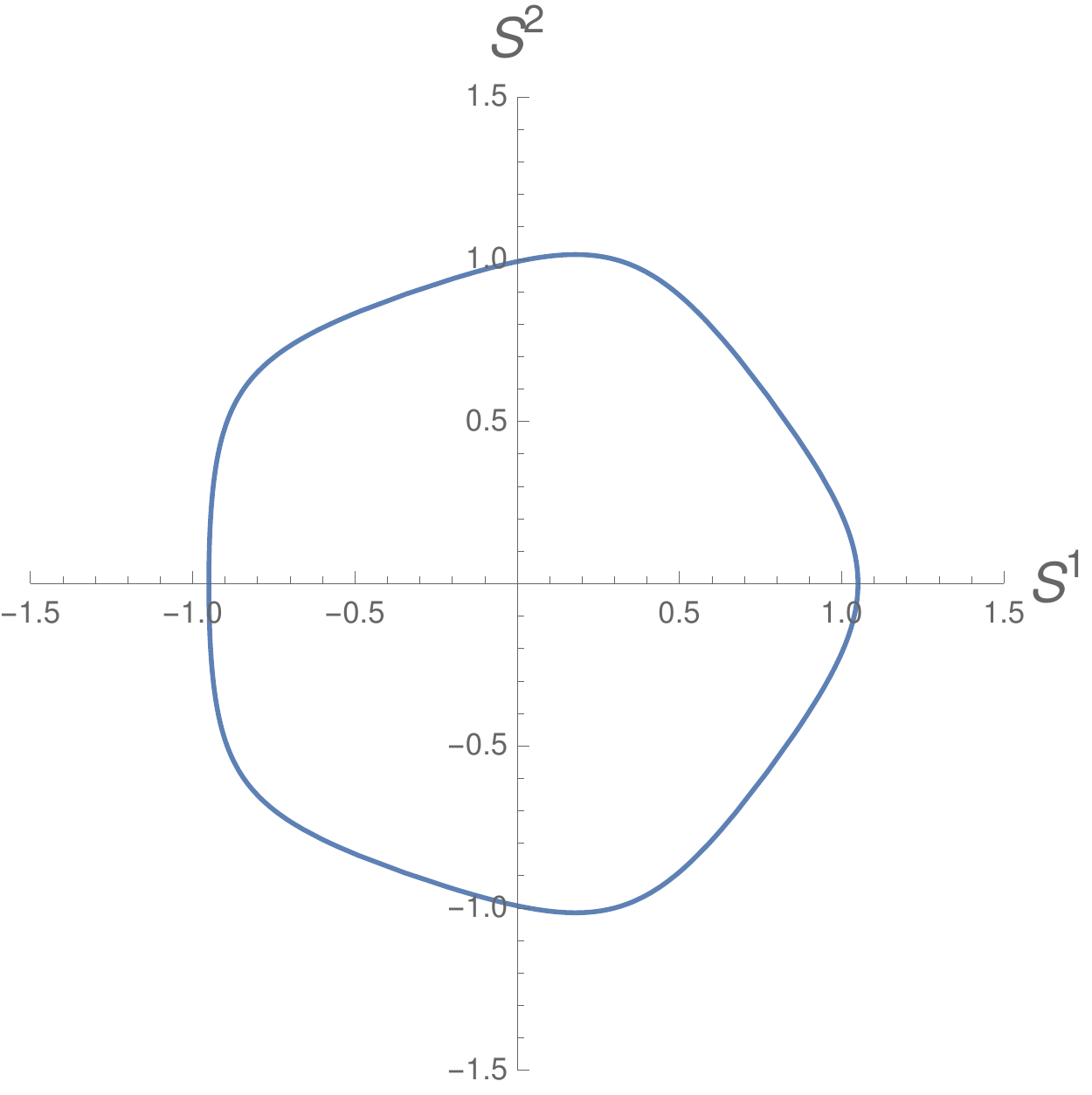}
\caption{$\frac{C_{11}}{2}=0.05$, $kz=0$, $\frac{\omega_g}{\omega}=3$, $t=T$.}
\label{BS4}
\end{figure}

\begin{figure}[h!]
\includegraphics[scale=0.50]{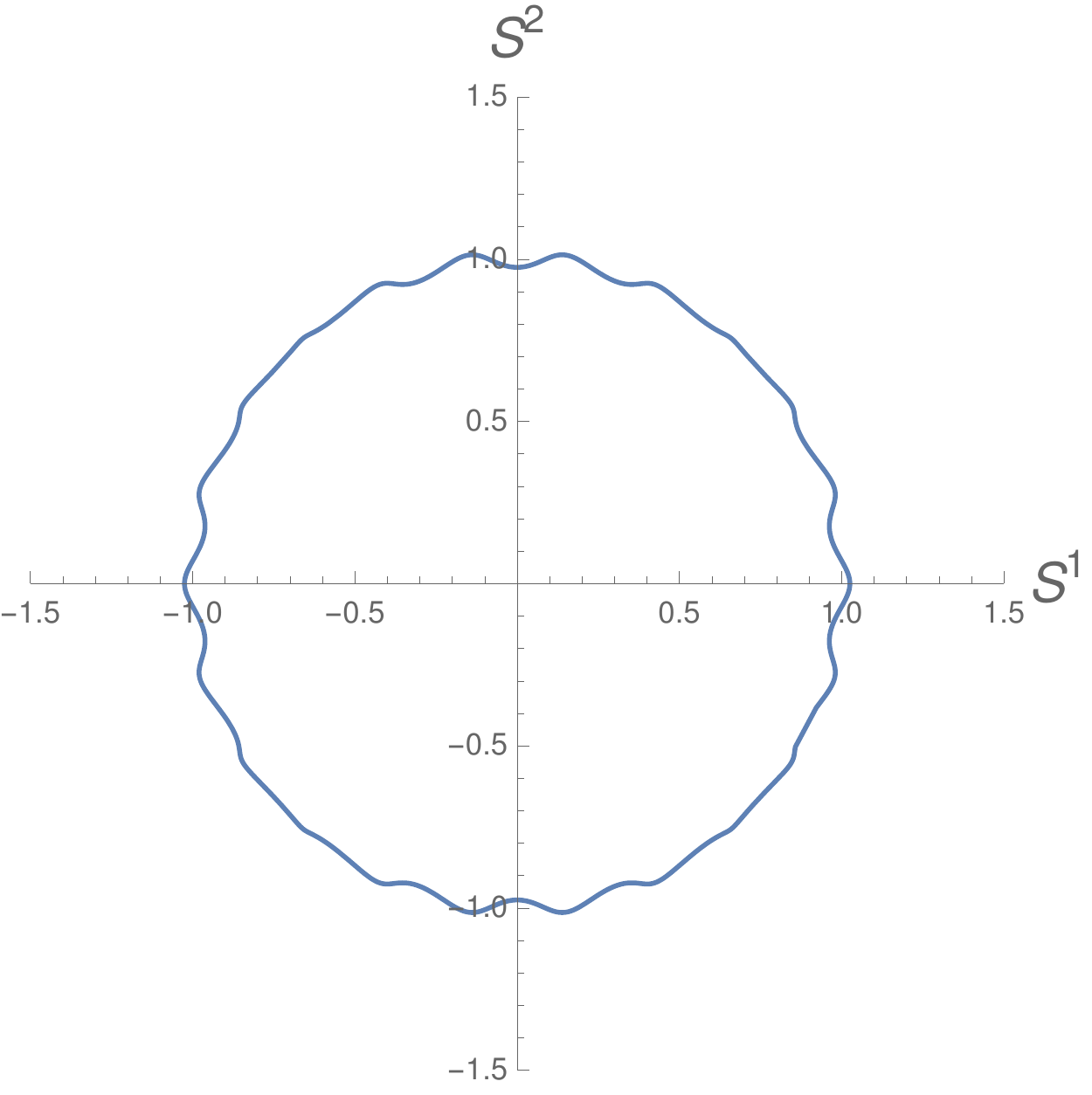}
\caption{$\frac{C_{11}}{2}=0.025$, $kz=0$, $\frac{\omega_g}{\omega}=20.$, $t=T$.}
\label{BS5}
\end{figure}

\begin{figure}[h!]
\includegraphics[scale=0.50]{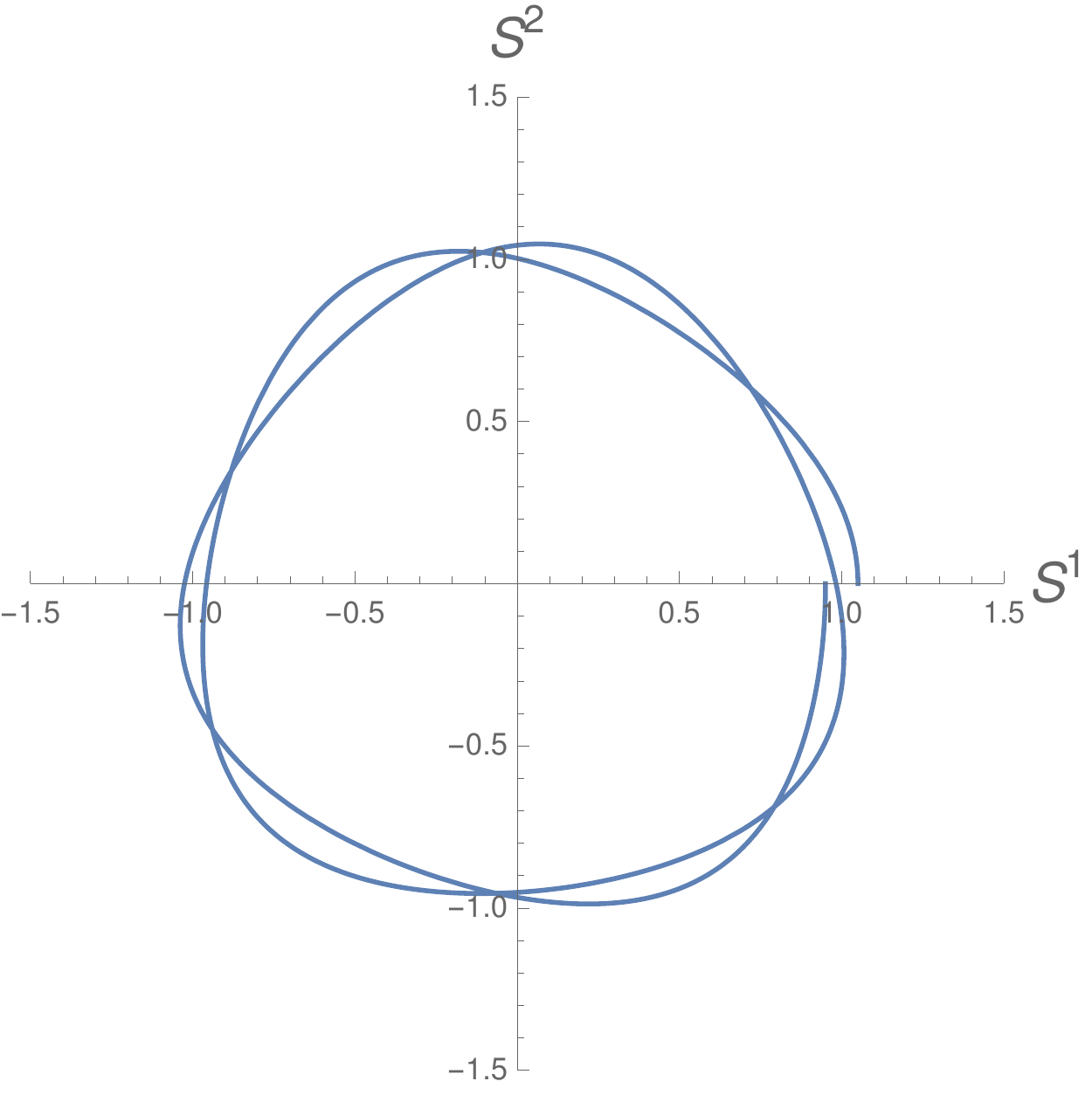}
\caption{$\frac{C_{11}}{2}=0.05$, $kz=0.$, $\frac{\omega_g}{\omega}=1.25$, $t=2~T$.}
\label{BS6}
\end{figure}

\begin{figure}[h!]
\includegraphics[scale=0.50]{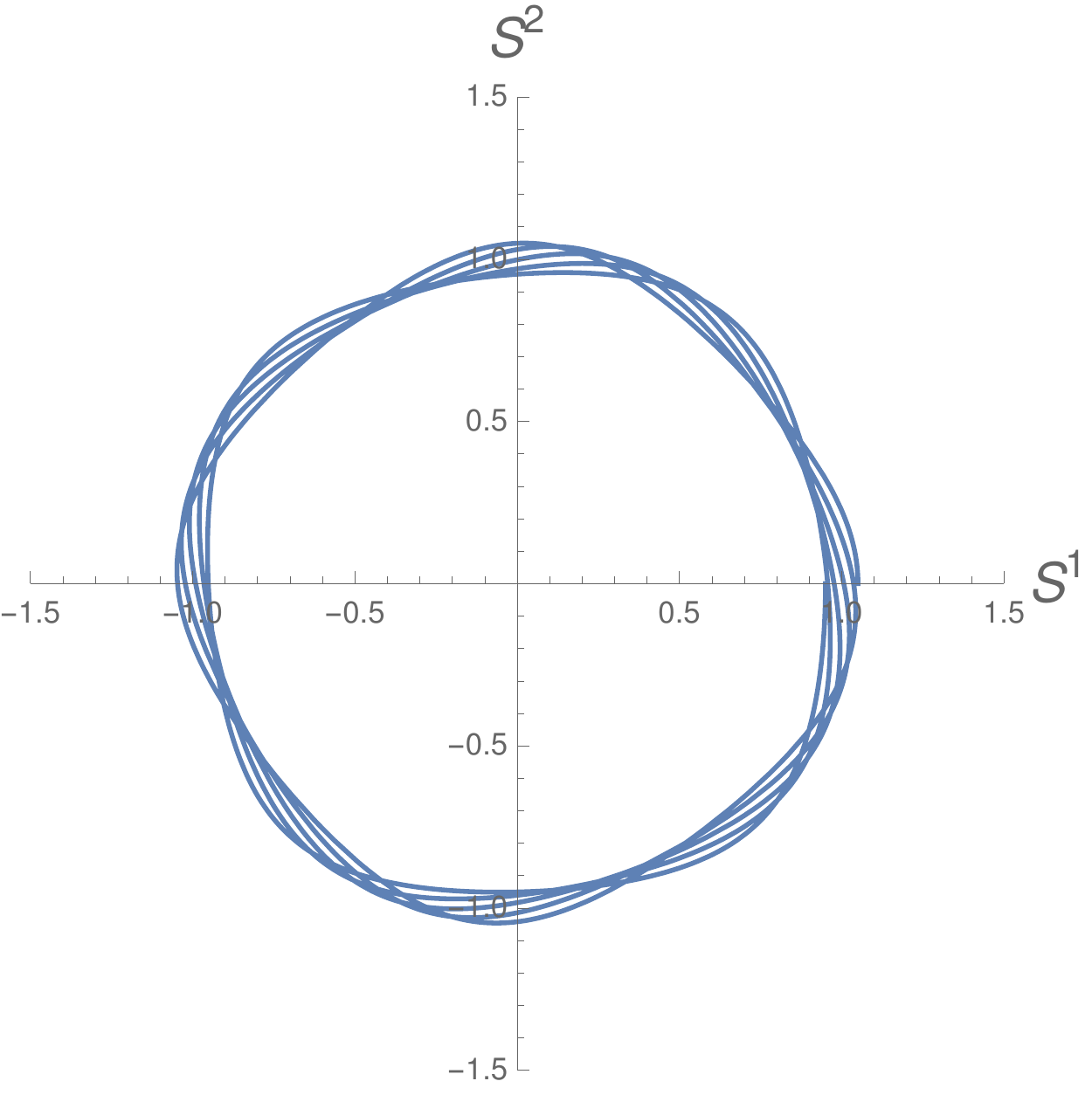}
\caption{$\frac{C_{11}}{2}=0.05$, $kz=0$, $\frac{\omega_g}{\omega}=2.1$, $t=5~T$.}
\label{BS7}
\end{figure}

\begin{figure}[h!]
\includegraphics[scale=0.50]{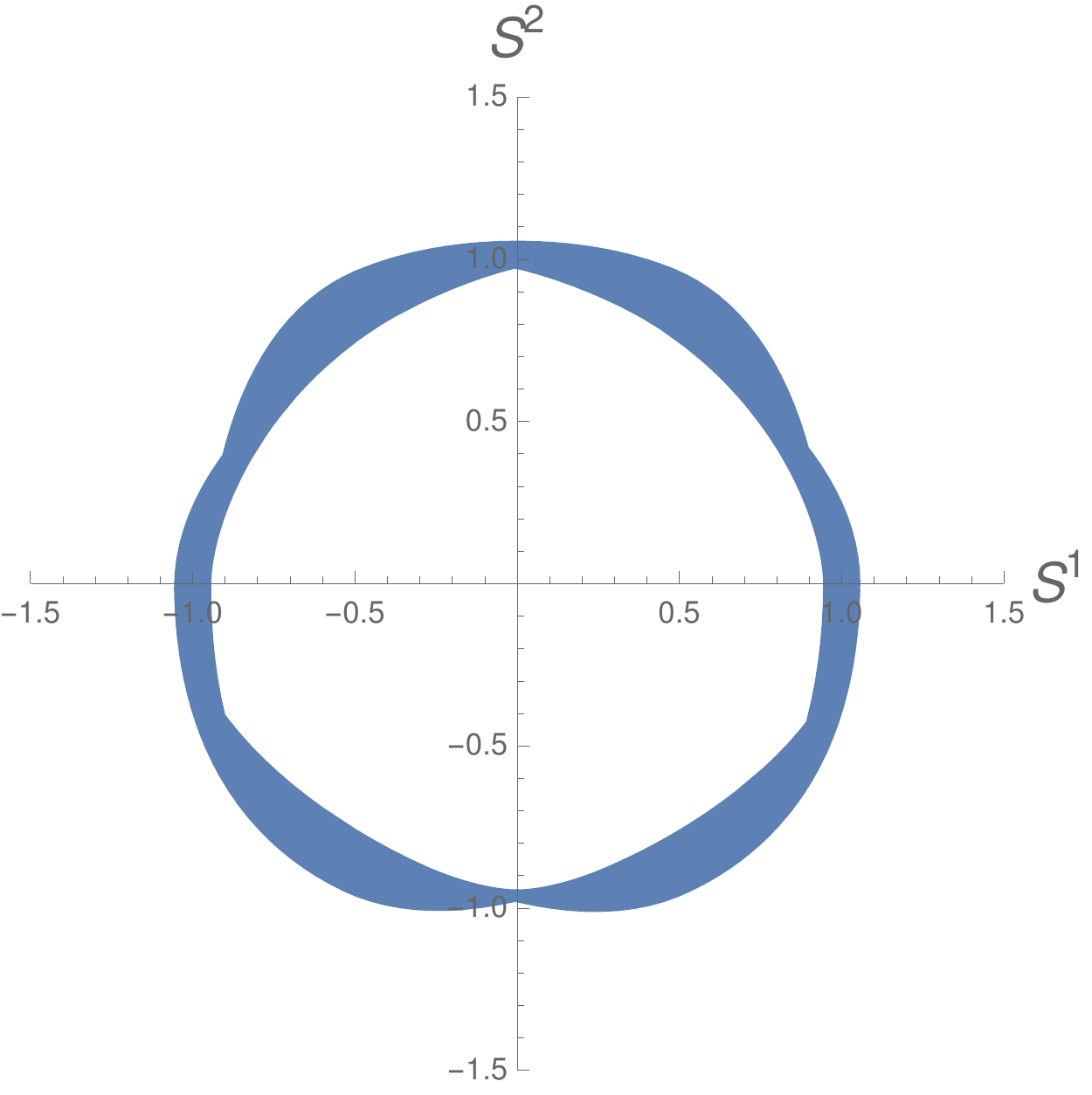}
\caption{$\frac{C_{11}}{2}=0.05$, $kz=0.$, $\frac{\omega_g}{\omega}=1.02$, $t=25~T$.}
 \label{BS8}
\end{figure}

\begin{figure}[h!]
\includegraphics[scale=0.50]{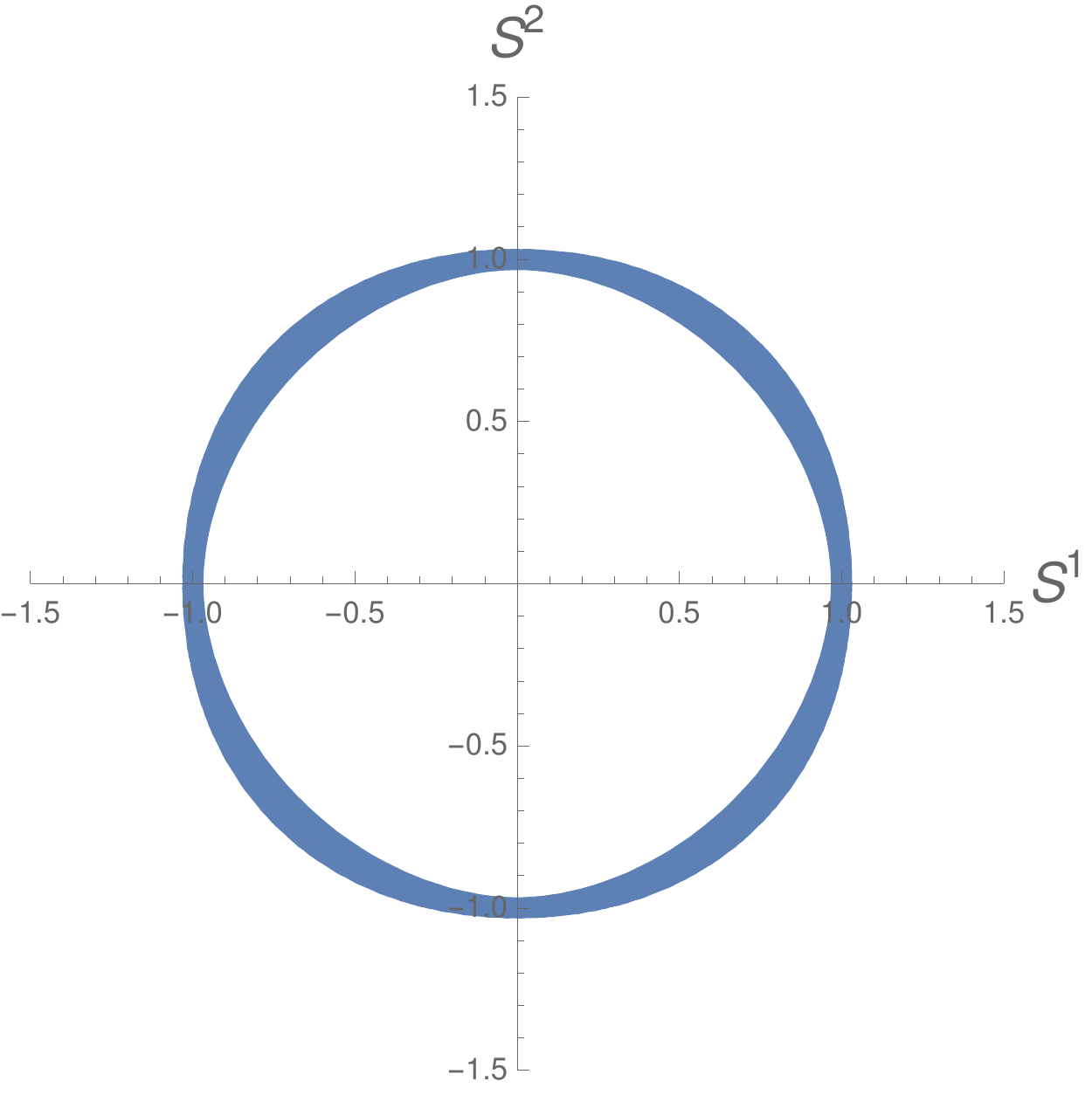}
\caption{$\frac{C_{11}}{2}=0.025$, $kz=0.$, $\frac{\omega_g}{\omega}=1.02$, $t=50~T$.}
\label{BS9}
\end{figure}

\begin{figure}[h]
\includegraphics[scale=0.50]{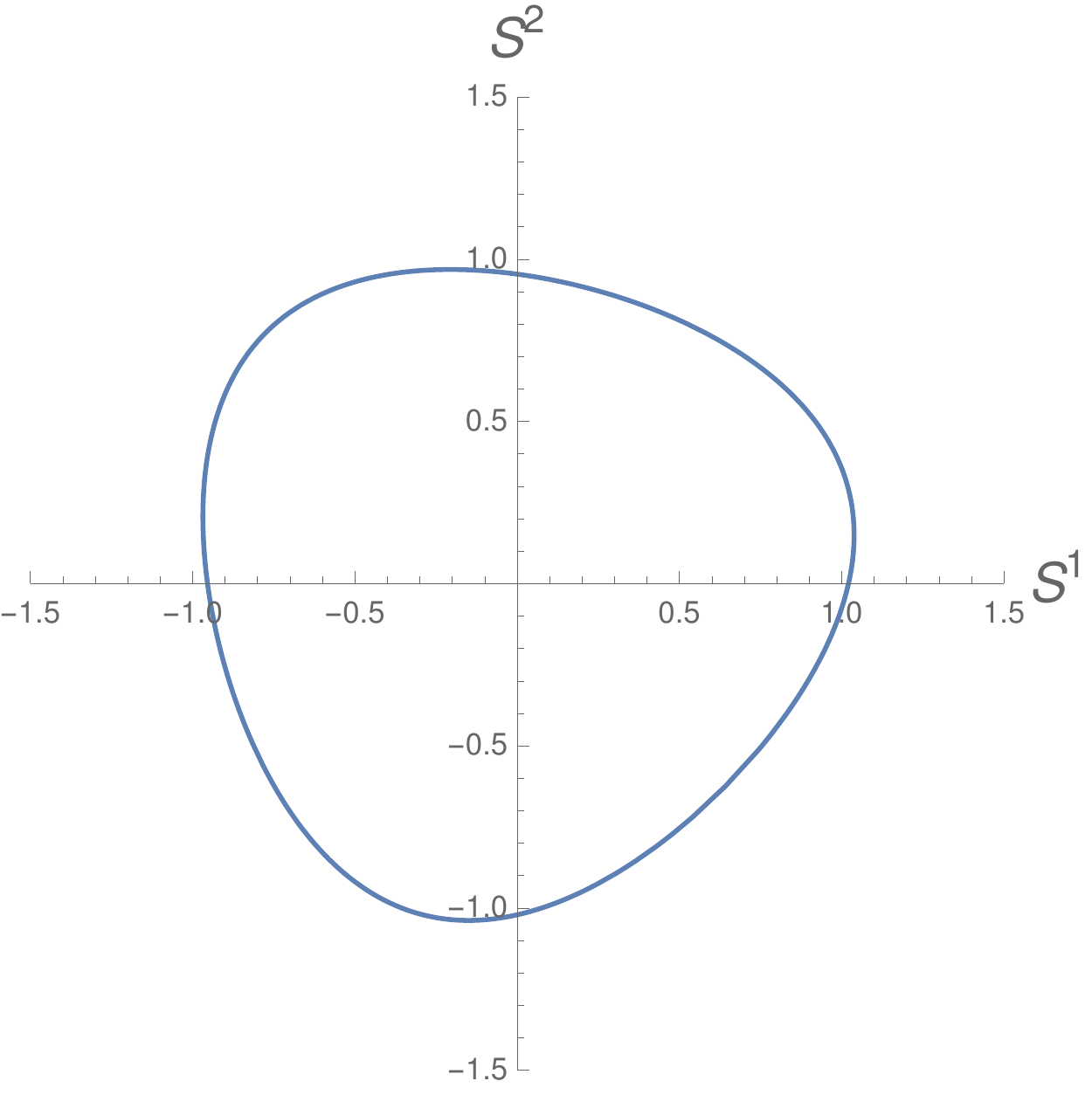}
\caption{$\frac{C_{11}}{2}=0.05$, $kz=\pi/4$, $\frac{\omega_g}{\omega}=1$, $t=T$.}
\label{BS10}
\end{figure}

\begin{figure}[h]
\includegraphics[scale=0.50]{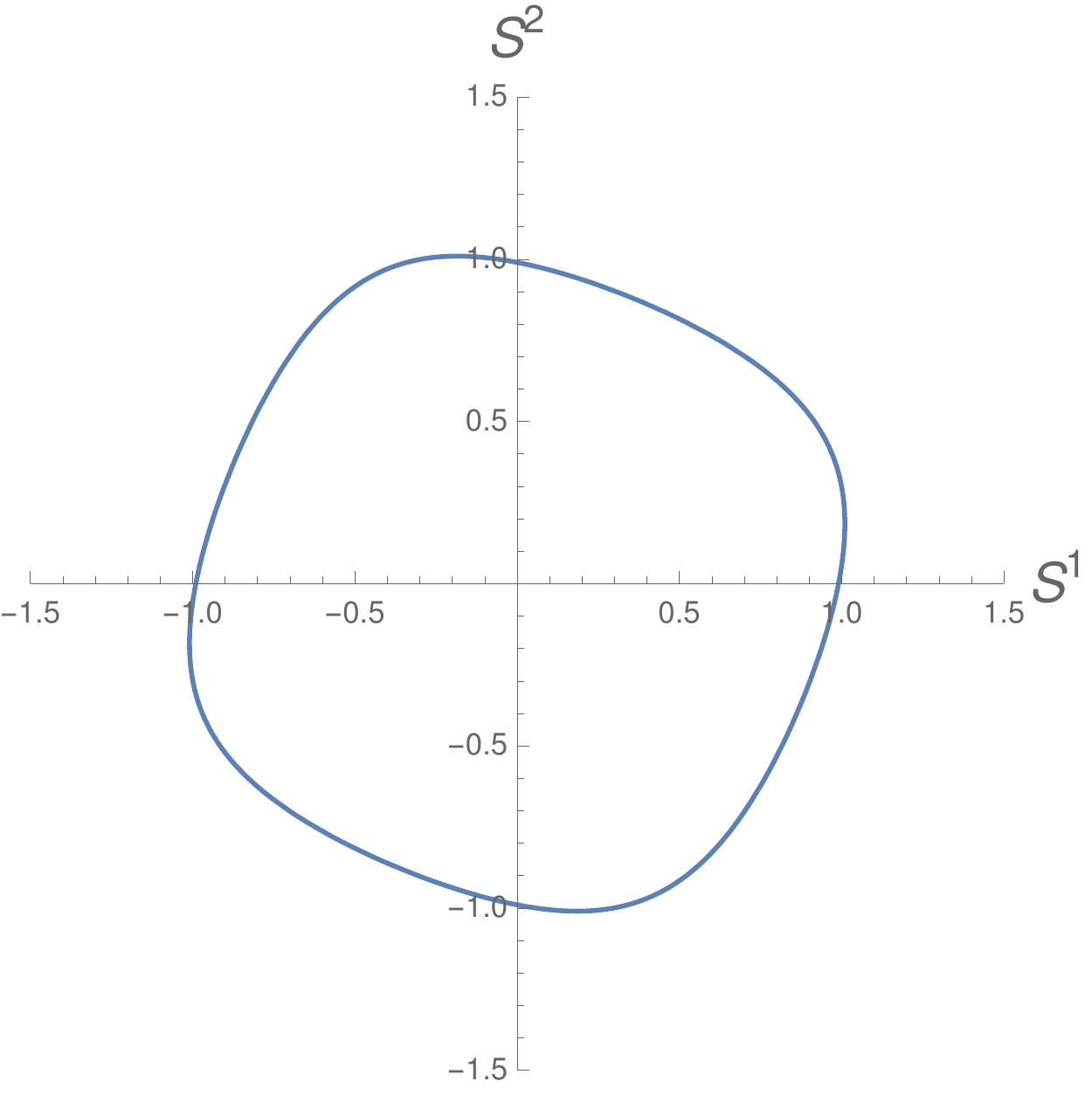}
\caption{$\frac{C_{11}}{2}=0.05$, $kz=\pi/4$ and $\frac{\omega_g}{\omega}=2$, $t=T$.}
\label{BS11}
\end{figure}

With the expressions~(\ref{ansatz1}) and (\ref{ansatz2})  for $S^\mu$ we are now able to display figures of the orbits, see Figs.~\ref{BS2}--\ref{BS11}. Since we want to compare the positions of the two particles at equal times we put $S_0=0$ in Eq.~(\ref{eta12}). To allow for better visibility we choose the amplitudes of the quadrupole oscillations in most diagrams unrealistically large, $C_{11}/2=0.05$. Only in Fig.~\ref{BS5} and Fig.~\ref{BS9} we choose slightly smaller values. In Figs.~\ref{BS2}-\ref{BS5} the orbits close after one revolution $T$ due to the integer frequency ratios $\frac{\omega_g}{\omega}$. With $\frac{\omega_g}{\omega}=1.25$ the path does not close yet after 2 revolutions in Fig.~\ref{BS6}, but it would after 4. In Fig.~\ref{BS7}, \ref{BS8} and \ref{BS9} we choose $\frac{\omega_g}{\omega}=2.1$ and $1.02$ respectively and need 10 and 50 revolutions to get a closed paths. In Figs.~\ref{BS8} and ~\ref{BS9} we compare with different values for the amplitude $C_{11}$. In the diagrams~\ref{BS10} and \ref{BS11} we modify the synchronisation of oscillation and rotational motion by choosing an additional phase, $kz=\pi/4$, leading to a rotation of the diagram compared to the figures~\ref{BS2} and ~\ref{BS3}.

These examples clearly reveal the influence of gravitational waves on circulating particles in the space-time structure of general relativity.

\section{\label{sec:D}Conclusion}
In this work we attempted to investigate the influence of gravitational waves on particles circulating in a central force field. We especially concentrated on periodic orbits and their perturbations.

After the introduction we recapitulated in chapter II the very well known approximate solution for slowly moving particles subjected to gravitational waves with frequency $\omega_g$. In chapter III we expanded  the problem to particles circulating with frequency $\omega$ in a central force field. To get a precision of the same order as for slowly moving test particles in refs.~\cite{Carroll,Cheng2005,schroeder2007,sharan2009,Braccini2016,Rebhan10} we had to take into account the centripetal force and to modify the time dependence of the azimuthal angle $\varphi=\omega t$ to $\phi$ of Eq.~(\ref{PolKoo}). Chapter IV is devoted to the graphical representation of the results. Besides the amplitude of the gravitational wave the ratio $\omega_g/\omega$ plays an important role. For integer values we get periodic orbits. For non-integer values of $\omega_g/\omega$ the paths are disturbed. For the shape of these perturbations we have to take into consideration that incident gravitational waves will have different polarisations and therefore different positions of the nodes. In any case a background of gravitational waves creates a kind of uncertainty in rotating systems.

The idea to this work is inspired by the silicon oil drop experiment of Yves Couder and his group. This is the only experiment, we know, which could give some idea why we can describe nature perfectly by quantum mechanics. A silicon droplet bouncing on a vibrating fluid bath creates waves interfering with the background field~\cite{Couder2006}. The resonant interaction of particle and field creates a wave field guiding moving droplets.

Thinking about the nature of a subquantum medium which could guide elementary particles we observe the importance of the Compton wave length which is related to the mass of particles. A natural type of background waves which could feel the mass of particles are gravitational waves. Waves which are not in resonance would lead to disturbances and to uncertainties in the position and momentum as we have seen in the investigation presented above. To get closer to quantum mechanics it would be necessary that the size of these displacements is related to the Compton wavelength. In the above linear treatment we find closed orbits for any integer ratio $\omega_g/\omega$. This does not agree with Bohr's quantisation condition. It would be interesting to take into account the non-linear terms of Einstein's equation and to investigate whether we get a relation to Bohr's quantisation condition.

\section*{Acknowledgments}
We thank Alexei A. Deriglazov and Jorge Daniel Casaleiro Lopes for valuable comments.
\section*{Conflict of Interests}
The authors declare that there are no conflicts of interest
regarding the publication of this paper.
\bibliographystyle{unsrt}
\bibliography{literatur}

\end{document}